\def\year{2019}\relax
\documentclass[sigconf]{acmart}

\pdfoutput=1
\usepackage{times}
\usepackage{helvet}

\usepackage{courier}
\usepackage[colorinlistoftodos,prependcaption,textsize=tiny]{todonotes}
\usepackage{url}  
\usepackage{graphicx}  
\usepackage{amsmath}
\usepackage{bbm}
\usepackage{booktabs} 
\usepackage{balance}
\usepackage[english]{babel}
\usepackage{fancyhdr,lipsum}
\usepackage[utf8]{inputenc}
\usepackage{algorithm}
\usepackage[noend]{algpseudocode}
\newcommand{\mathds}[1]{\text{\usefont{U}{dsrom}{m}{n}#1}}

\newcommand{\argmax}[1]{\underset{#1}{\operatorname{arg}\,\operatorname{max}}\;}

\copyrightyear{2019} 
\acmYear{2019} 
\setcopyright{rightsretained} 
\acmConference[FLAIRS '19]{AAAI Conference on AI, Ethics, and Society}{May 18--22, 2019}{Sarasota, Florida, USA}
\acmBooktitle{AAAI Florida Artificial Intelligence Research Society  (FLAIRS '19), May 18--22, 2019, Sarasota, Florida, USA}
\acmDOI{10.1145/XXXXXXX}
\acmISBN{978-1-4503-XXXXXX}
 
\begin{document}
%
\title{Managing Popularity Bias in Recommender Systems \\
with Personalized Re-ranking}

\author{Himan Abdollahpouri}
\email{himan.abdollahpouri@colorado.edu}
\affiliation{%
  \institution{Department of Information Science \\ University of Colorado Boulder}
}

\author{Robin Burke}
\email{robin.burke@colorado.edu}
\affiliation{%
  \institution{Department of Information Science \\ University of Colorado Boulder}
}
\author{Bamshad Mobasher}
\email{mobasher@cs.depaul.edu}
\affiliation{%
  \institution{Web Intelligence Lab, DePaul University}
}

\begin{abstract}

Many recommender systems suffer from popularity bias: popular items are recommended frequently while less popular, niche products, are recommended rarely or not at all. However, recommending the ignored products in the ``long tail'' is critical for businesses as they are less likely to be discovered. In this paper, we introduce a personalized diversification re-ranking approach to increase the representation of less popular items in recommendations while maintaining acceptable recommendation accuracy. Our approach is a post-processing step that can be applied to the output of any recommender system. We show that our approach is capable of managing popularity bias more effectively, compared with an existing method based on regularization. We also examine both new and existing metrics to measure the coverage of long-tail items in the recommendation.
\end{abstract}
\maketitle
\section{Introduction}
\noindent 

Recommender systems have an important role in e-commerce and information sites, helping users find new items. One obstacle to the effectiveness of recommenders is the problem of popularity bias \cite{bellogin2017statistical}: collaborative filtering recommenders typically emphasize popular items (those with more ratings) over other ``long-tail'' items~\cite{longtailrecsys} that may only be popular among small groups of users. Although popular items are often good recommendations, they are also likely to be well-known. So delivering only popular items will not enhance new item discovery and will ignore the interests of users with niche tastes. It also may be unfair to the producers of less popular or newer items since they are rated by fewer users. 

\begin{figure}[h]
    \centering
    \includegraphics[width=2.5in]{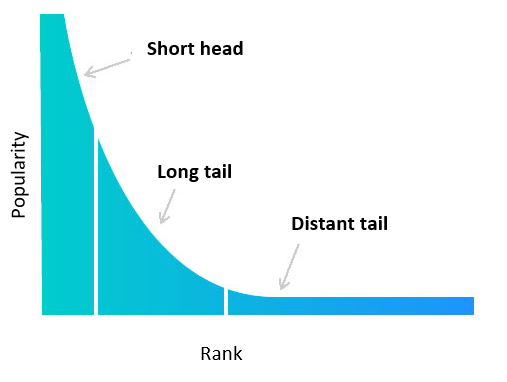}
    \caption{The long-tail of item popularity. }
    \label{fig:long-tail}
\end{figure}

Figure~\ref{fig:long-tail} illustrates the long-tail phenomenon in recommender systems. The $y$ axis represents the number of ratings per item and the $x$ axis shows the product rank. The first vertical line separates the top 20\% of items by popularity -- these items cumulatively have many more ratings than the 80\%  tail items to the right. These ``short head'' items are the very popular items, such as blockbuster movies in a movie recommender system, that garner much more viewer attention. Similar distributions can be found in other consumer domains. 

The second vertical line divides the tail of the distribution into two parts. We call the first part the \textit{long tail}: these items are amenable to collaborative recommendation, even though many algorithms fail to include them in recommendation lists. The second part, the \textit{distant tail}, are items that receive so few ratings that meaningful cross-user comparison of their ratings becomes unreliable. For these cold-start items, content-based and hybrid recommendation techniques must be employed. Our work in this paper is concerned with collaborative recommendation and therefore focuses on the long-tail segment.

We present a general and flexible approach for controlling the balance of item exposure in different portions of the item catalog as a post-processing phase for standard recommendation algorithms. Our work is inspired by \cite{santos2010exploiting} where authors introduced a novel probabilistic framework called \textit{xQuAD} for Web search result
diversification which aims to generate search results that explicitly account for various aspects associated with an under-specified query. We adapt the xQuAD approach to the popularity bias problem. Our approach enables the system designer to tune the system to achieve the desired trade-off between accuracy and better coverage of long-tail, less popular items. 

\subsection{Related Work}
Recommending serendipitous items from the long tail is generally considered to be a key function of recommendation \cite{anderson2006long}, as these are items that users are less likely to know about. Authors in ~\cite{brynjolfsson2006niches} showed that 30-40\% of Amazon book sales are represented by titles that would not normally be found in brick-and-mortar stores. 

Long-tail items are also important for generating a fuller understanding of users' preferences. Systems that use active learning to explore each user's profile will typically need to present more long tail items because these are the ones that the user is less likely to know about, and where user's preferences are more likely to be diverse~\cite{resnick2013bursting}. 

Finally, long-tail recommendation can also be understood as a social good. A market that suffers from popularity bias will lack opportunities to discover more obscure products and will be, by definition, dominated by a few large brands or well-known artists~\cite{celma2008hits}. Such a market will be more homogeneous and offer fewer opportunities for innovation and creativity.

The idea of the long-tail of item popularity and its impact on recommendation quality has been explored by some researchers \cite{brynjolfsson2006niches,longtailrecsys}. In those works, authors tried to improve the performance of the recommender system in terms of accuracy and precision, given the long-tail in the ratings. Our work, instead, focuses on reducing popularity bias and balancing the representation of items across the popularity distribution.  

A regularization-based approach to improving long tail recommendations is found in \cite{abdollahpouri2017controlling}. One limitation with that work is that this work is restricted to factorization models where the long-tail preference can be encoded in terms of the latent factors. In contrast, a re-ranking approach can be applied to the output of any algorithm. Another limitation of that work is that it does not account for user tolerance towards long-tail items: the fact that there may be some users only interested in popular items. In our model, we take personalization of long-tail promotion into account as well. 

And finally, there is substantial research in recommendation diversity, where the goal is to avoid recommending too many similar items~\cite{zhou2010solving,castells2011novelty,zhang2008avoiding}. Personalized diversity is also another related area of research where the amount of diversification is dependent on the user's tolerance for diversity \cite{eskandanian2017clustering,wasilewski2018intent}. Another similar work to ours is \cite{vargas2012explicit} where authors used a modified version of xQuAD called relevance based xQuAD for intent-oriented diversification of search results and recommendations. Another work used a similar approach but for fairness-aware recommendation \cite{liu2018personalizing} where xQuAD was used to make a fair representation of items from different item providers. Our work is different from all these previous diversification approaches in that it is not dependent on the characteristics of items, but rather on the relative popularity of items. 

\section{Controlling Popularity Bias}

\subsection{xQuAD}
Result diversification has been studied in the context of information retrieval, especially for web search engines, which have a similar goal to find a ranking of documents that together provide a complete
coverage of the aspects underlying a query \cite{santos2015search}. EXplicit Query
Aspect Diversification (xQuAD) \cite{santos2010exploiting} explicitly accounts for the various aspects associated with an under-specified query. Items are selected iteratively by estimating how well a given document satisfies an uncovered aspect. 

In adapting this approach, we seek to recognize the difference among users in their interest in long-tail items. Uniformly-increasing diversity of items with different popularity levels in the recommendation lists may work poorly for some users. We propose a variant that adds a personalized bonus to the items that belong to the under-represented group (i.e. the long-tail items). The personalization factor is determined based on each user's historical interest in long-tail items.

\section{Methodology}

We build on the \textit{xQuAD} model to control popularity bias in recommendation
algorithms. We assume that for a given user $u$, a ranked recommendation list $R$ has already been generated by a base recommendation algorithm. The task of the modified xQuAD method is to produce a new re-ranked list $S$ ($|S|<|R|$) that manages popularity bias while still being accurate. 

The new list is built iteratively according to the following criterion:
\begin{equation}\label{eq:1}
    P(v|u)+\lambda P(v,S'|u)
\end{equation}
where $P(v|u)$ is the likelihood of user $u \in U$ being interested in
item $v \in V$, independent of the items on the list so far as, predicted by the base recommender. The second term $P(v, S'|u)$ denotes the likelihood of user u being interested in an item $v$ as an item not in the currently generated list $S$.

Intuitively, the first term
incorporates ranking accuracy while the second term promotes
diversity between two different categories of items (i.e. short head and long tail). The parameter $\lambda$ controls how strongly controlling popularity bias is weighted in general. The item
that scores most highly under the equation \ref{eq:1} is added to the
output list $S$ and the process is repeated until $S$ has achieved the
desired length.

To achieve more diverse recommendation containing items from both short head and long tail items, the marginal likelihood $P(v, S'|u)$ over both
item categories long-tail head ($\Gamma$) and short head ($\Gamma$') is computed by:

\begin{equation}\label{eq:2}
    P(v,S'|u)=\sum_{d \in \{\Gamma , \Gamma '\}}P(v,S'|d)P(d|u)
\end{equation}

Following the approach of \cite{santos2010exploiting}, we assume that the remaining
items are independent of the current contents of $S$ and that the items
are independent of each other given the short head and long tail categories. Under these assumptions, we can compute $P(v, S'|d)$ in Eq.\ref{eq:2} as

\begin{equation} \label{eq:3}
    P(v,S'|d)=P(v|d)P(S'|d)=
    P(v|d)\prod_{i \in S} (1-P(i|d,S))
\end{equation}

By substituting equation \ref{eq:3} into equation \ref{eq:2}, we can obtain
\begin{equation} \label{eq:4}
     score=(1-\lambda) P(v |u)+\lambda\sum_{c \in \{\Gamma , \Gamma '\}}P(c|u)P(v|c)\prod_{i \in S}(1-P(i|c,S))
\end{equation}
where $P(v|d)$ is equal to 1 if $v \in d$ and 0 otherwise. 

We measure $P(i|d,S)$ in two different ways to produce two different algorithms. The first way is to use the same function as $P(v|d)$, an indicator function where it equals to 1 when item $i$ in list $S$ already covers category $d$ and 0 otherwise. We call this method \textit{Binary xQuAD} and it is how original xQuAD was introduced. Another method that we present in this paper is to find the ratio of items in list $S$ that covers category $d$. We call this method \textit{Smooth xQuAD}.

The likelihood $P(d|u)$ is the measure of user preference over different item categories. In other words, it measures how much each user is interested in short head items versus long tail items. We calculate this likelihood by the ratio of items in the user profile which belong to category $d$. 

In order to select the next item to add to $S$, we compute a re-ranking score for each item in $R \backslash S$ according to Eq. \ref{eq:4}. For an item $v' \in d$, if $S$ does not cover $d$, then an additional positive term will be added to the estimated user preference $P(v'|u)$. Therefore, the chance that it will be selected is larger, balancing accuracy and popularity bias. 

In Binary xQuAD, the product term $\prod_{i \in S}(1-P(i|d,S))$ is only equal to 1 if the current items
in $S$ have not covered the category $d$ yet. Binary xQuAD is, therefore, optimizing for a \textit{minimal} re-ranking of the original list by including the best long-tail item it can, but not seeking diversity beyond that.

\section{Experiment}
In this section, we test our proposed algorithm on two public datasets. The first is the well-known Movielens 1M dataset that contains 1,000,209 anonymous ratings of approximately 3,900 movies made by 6,040 MovieLens users~\cite{movielens}. The second dataset is the Epinions dataset, which is gathered from a consumers opinion site where users can review items \cite{massa2007trust}. This dataset has the total number of 664,824 ratings given by 40,163 users to 139,736 items. In Movielens, each user has a minimum of 20 ratings but in Epinions, there are many users with only a single rated item. 

Following the data reduction procedure in \cite{abdollahpouri2017controlling}, we removed users who had fewer than 20 ratings from the Epinion dataset, as users with longer profiles are much more likely to have long-tail items in their profiles. MovieLens dataset already consists of only users with more than 20 ratings. The retained users were those likely to have rated enough long-tail items so that our objective could be evaluated in a train / test scenario. We also removed distant long-tail items from each dataset using a limit of 20 ratings, a number 20 is chosen to be consistent with the cut-off for users.

After filtering, the MovieLens dataset has 6,040 users who rated 3043 movies with a total number of 995,492 ratings, a reduction of about 0.4\%. Applying the same criteria to the Epinions dataset decreases the data to 220,117 ratings given by 8,144 users to 5,195 items, a reduction of around 66\%. We split the items in both datasets into two categories: long-tail ($\Gamma$) and short head ($\Gamma$')in a way that short head items correspond to \%80 of the ratings while long-tail items have the rest of the \%20 of the ratings. We plan to consider other divisions of the popularity distribution in future work. For MovieLens, the short-head items were those with
more than 506 ratings. In Epinions, a short-head item needed only to have more than 73 ratings.

\section{Evaluation}
The experiments compare four algorithms. Since we are concerned with ranking performance, we chose as our baseline algorithm RankALS, a pair-wise learning-to-rank algorithm. We also include the regularized long-tail diversification algorithm in \cite{abdollahpouri2017controlling} (indicated as \textit{LT-Reg} in the figures.) We used the output from RankALS as input for the two re-ranking variants described above: Binary xQuAD and Smooth xQuAD, marked \textit{Binary} and \textit{Smooth} in the figures. We compute lists of length 100 from RankALS and pass these to the re-ranking algorithms to compute the final list of 10 recommendations for each user. We used the implementation of RankALS in LibRec 2.0\footnote{www.librec.net} for all experiments.

In order to evaluate the effectiveness of algorithms in mitigating popularity bias we use four different metrics:

\textbf{Average Recommendation Popularity (ARP)}: This measure from \cite{yin2012challenging} calculates the average popularity of the recommended items in each list. For any given recommended item in the list, we measure the average number of ratings for those items. More formally:
\begin{equation}\label{eq:arp}
     ARP=\frac{1}{|U_{t}|}\sum_{u \in U_{t}} \frac{\sum_{i \in L_u}\phi(i)}{|L_u|}
\end{equation}
where $\phi(i)$ is the number of times item $i$ has been rated in the training set. $L_u$ is the recommended list of items for user $u$ and $|U_{t}|$ is the number of users in the test set.

\textbf{Average Percentage of Long Tail Items (APLT)}: As used in \cite{abdollahpouri2017controlling}, this metric measures the average percentage of long tail items in the recommended lists and it is defined as follows:
\begin{equation}\label{eq:aplt}
     APLT=\frac{1}{|U_{t}|}\sum_{u \in U_{t}} \frac{|\{i, i \in (L_u \cap \Gamma)\} |}{|L_u|}
\end{equation}
 This measure gives us the average percentage of items in users' recommendation lists that belong to the long tail set.

\textbf{Average Coverage of Long Tail items (ACLT)}: We introduce another metric to evaluate how much exposure long-tail items get in the entire recommendation. One problem with $APLT$ is that it could be high even if all users get the same set of long tail items. $ACLT$ measures what fraction of the long-tail items the recommender has covered: 
\begin{equation}\label{eq:aclt}
    ACLT=\frac{1}{|U_{t}|}\sum_{u \in U_{t}} \sum_{i \in L_u} \mathds{1}({i \in \Gamma})
\end{equation}
where $\mathds{1}({i \in \Gamma})$ is an indicator function and it equals to 1 when i is in $\Gamma$. This function is related to the \textit{Aggregate Diversity} metric of \cite{adomavicius2012improving} but it looks only at the long-tail part of the item catalog.

In addition to the aforementioned long tail diversity metrics, we also evaluate the accuracy of the ranking algorithms in order to examine the diversity-accuracy trade-offs. For this purpose we use the standard \textit{Normalized Discounted cumulative Gain} (\textit{NDCG}) measure of ranking accuracy. 

\begin{figure*}[t]
    \centering
    \includegraphics[width=6in]{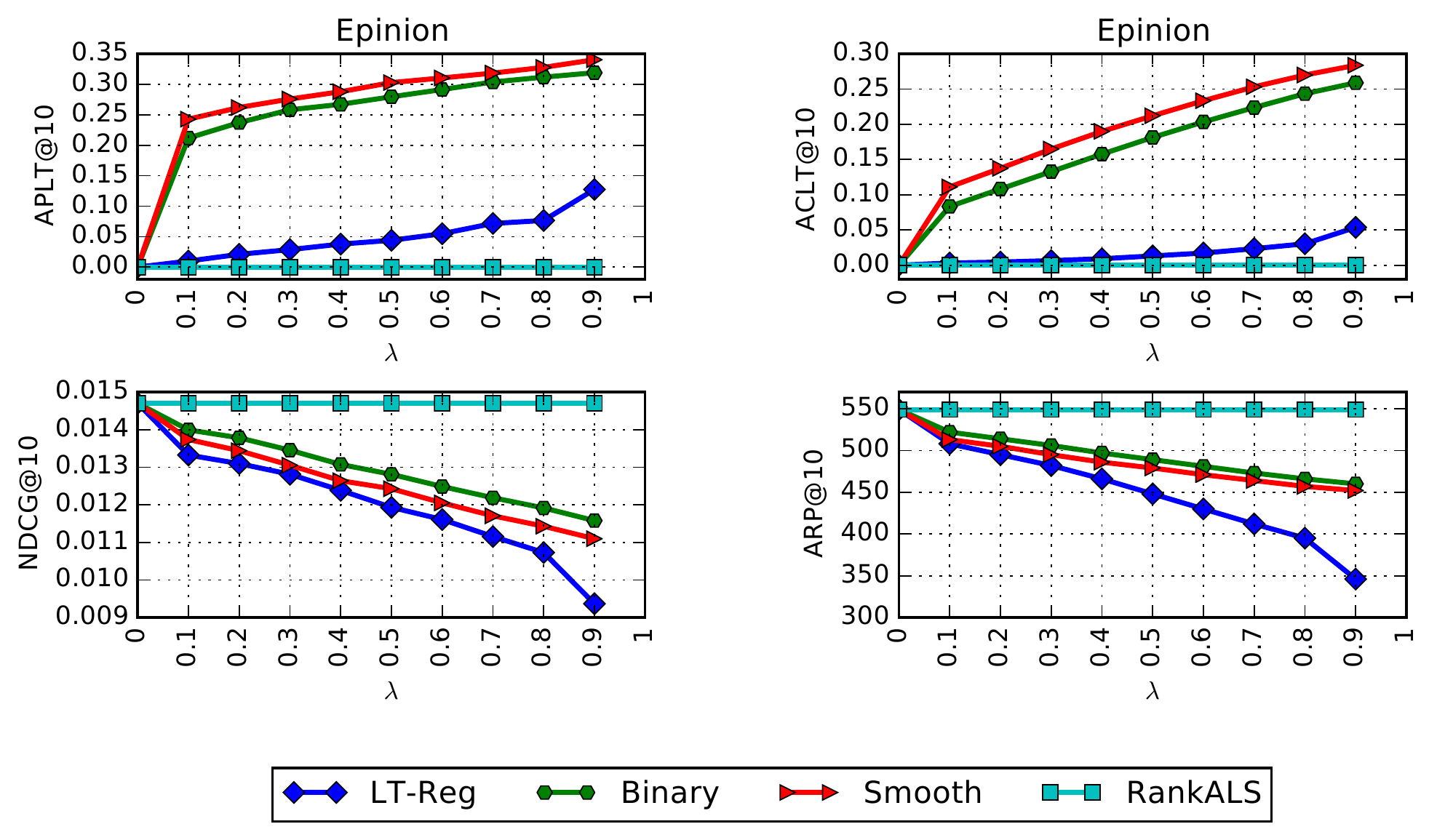}
    \caption{Results for the Epinions dataset}
    \label{fig:ep-results1}
\end{figure*}
\begin{figure*}[t]
    \centering
    \includegraphics[width=6in]{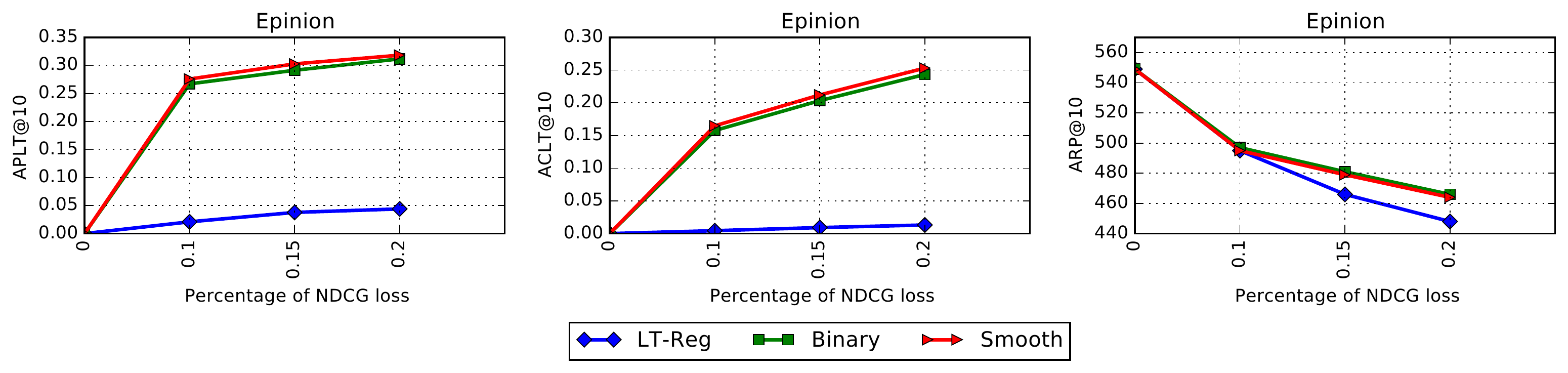}
    \caption{Comparison of popularity bias control for different algorithms at different levels of NDCG loss (Epinions)}
    \label{fig:ep-results2}
\end{figure*}
\section{Results}

Figure~\ref{fig:ep-results1} shows the results for the Epinions dataset across the different algorithms using a range of values for $\lambda$. (Note that the LT-Reg algorithm uses the parameter $\lambda$ to control the weight placed on the long-tail regularization term.) All results are averages from five-fold cross-validation using a \%80 -\%20 split for train and test, respectively. As expected, the diversity scores improve for all algorithms, with some loss of ranking accuracy. Differences between the algorithms are evident, however. The exposure metric (ACLT) plot shows that the two re-ranking algorithms, and especially the Smooth version, are doing a much better job of exposing items across the long-tail inventory than the regularization method. The ranking accuracy shows that, as expected, the Binary version does slightly better as it performs minimal adjustment to the ranked lists. LT-Reg is not as effective at promoting long-tail items, either by the list-wise APLT measure or by the catalog-wise ACLT.

\begin{figure*}[t]
    \centering
    \includegraphics[width=6in]{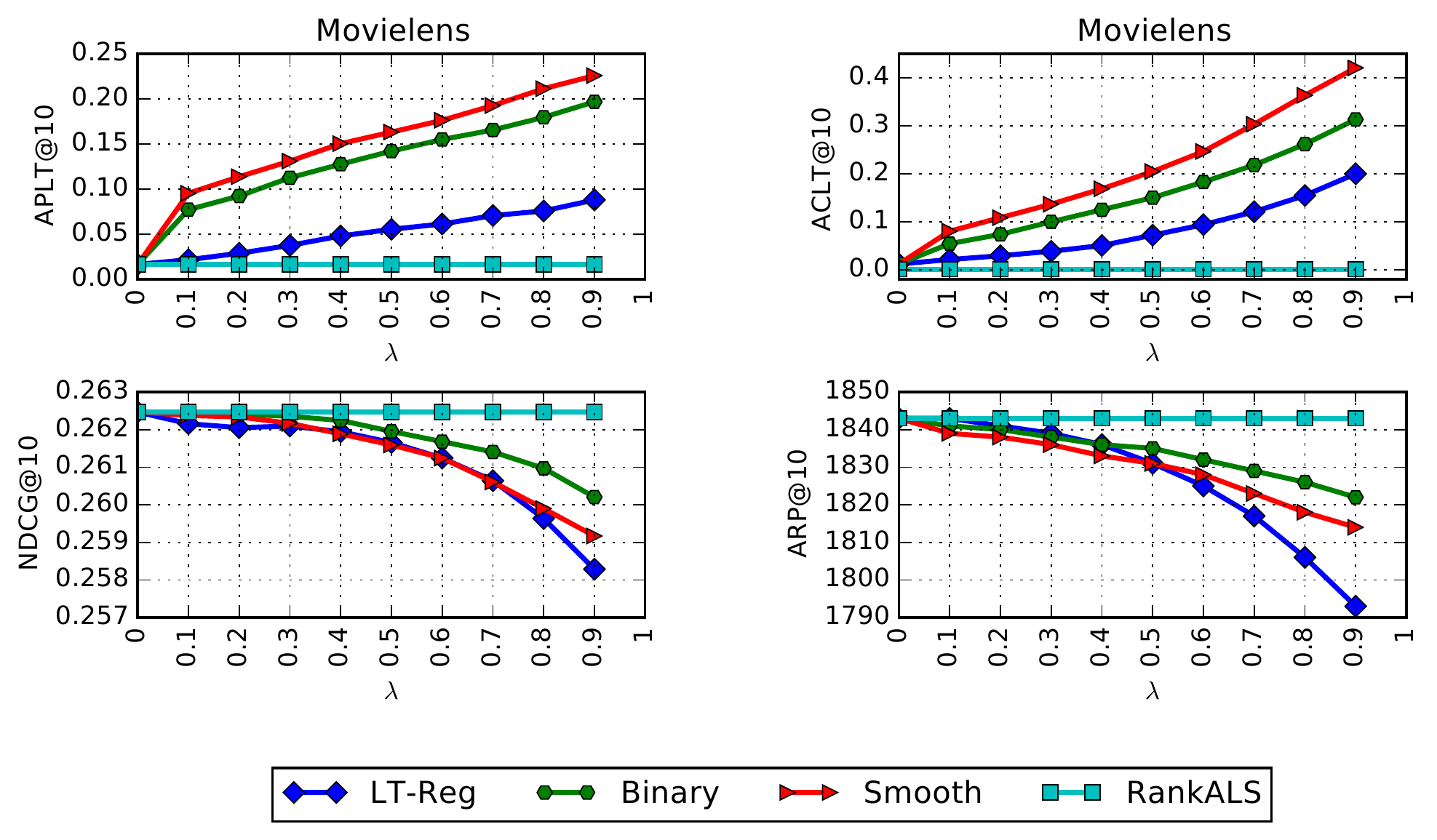}
    \caption{Results for the MovieLens dataset}
    \label{fig:ml-results1}
\end{figure*}

\begin{figure*}[t]
    \centering
    \includegraphics[width=6in]{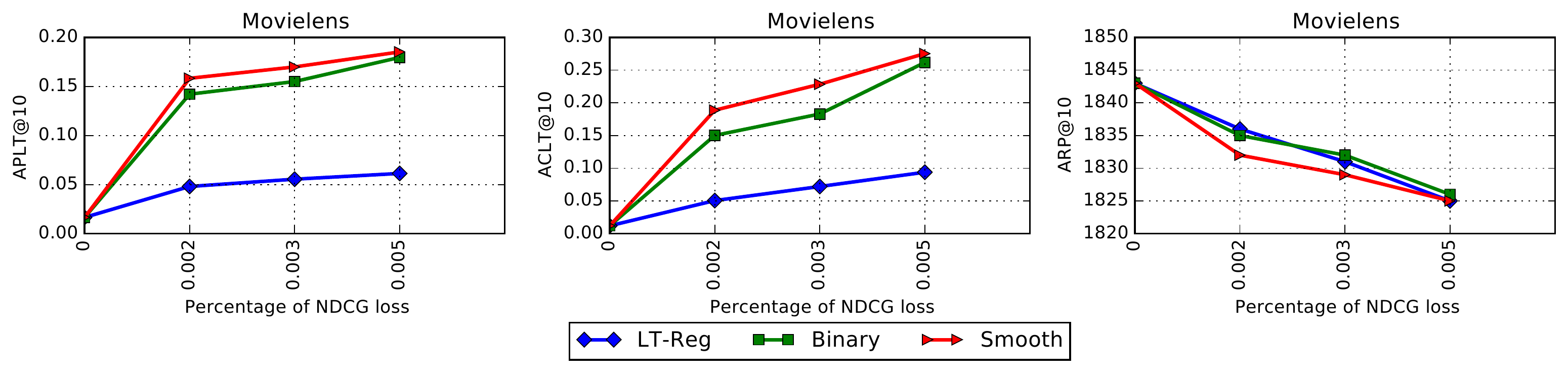}
    \caption{Comparison of popularity bias control for different algorithms at different levels of NDCG loss (MovieLens)}
    \label{fig:ml-results2}
\end{figure*}

Another view of the same results is provided in Figure~\ref{fig:ep-results2}. Here we look at the long-tail diversity metrics relative to NDCG loss, which clarifies the patterns seen in Figure~\ref{fig:ep-results1}. We see that the Binary and Smooth algorithms are fairly similar in terms of diversity-accuracy trade-off, while LT-Reg has a distinctly lower and flatter improvement curve with increased loss of ranking accuracy. ARP metric is the only one where the algorithms are fairly similar, especially at lower values of NDCG loss.

The MovieLens dataset shows different relative performance across the algorithms as seen in Figure~\ref{fig:ml-results1}. The Smooth re-ranking method shows a more distinct benefit and LT-Reg is somewhat more effective. This finding is confirmed in the relative results shows in Figure~\ref{fig:ml-results2}, which also shows the algorithms having quite similar values for the ARP metric, in spite of the differences on the other metrics.

Comparing the two datasets, we see that long-tail diversification is more of a challenge in the sparser Epinions dataset. With 10\% NDCG loss, it is possible to bring exposure to around 15\% of the long-tail catalog in Epinions; whereas for MovieLens, 0.2\% loss yields an equivalent or greater benefit. LT-Reg is much less effective. (In both datasets, the baseline value is very close to zero.) The average number of long-tail items in each recommendation list shows a similar pattern. 

In the sparser dataset, the Binary and Smooth measures are similar in performance, but differences appear in MovieLens, where the Smooth algorithm shows stronger improvement in the ACLT measure, particularly. This effect is most likely due to the fact that in the sparser data, it is more difficult to find a single long-tail item to promote into a recommendation list, with greater accuracy cost in doing so. In MovieLens, these higher-quality items appear more often and the Smooth objective values the promotion of multiple such items into the recommendation lists.

Another conclusion we can draw is that the ARP measure is not a good measure of long-tail diversity when it is used only on its own. It has the benefit of not requiring the experimenter to set a threshold distinguishing long-tail and short-head items. However, as we see here, algorithms can have very similar ARP performance and be quite different in terms of their ability to cover the long-tail catalog and to promote long-tail items to users. So it is important to look at all these metrics together.


\section{Conclusion and future work}
Adequate coverage of long-tail items is an important factor in the practical success of business-to-consumer organizations and information providers. Since short-head items are likely to be well known to many users, the ability to recommend items outside of this band of popularity will determine if a recommender system can introduce users to new products and experiences. Yet, it is well-known that recommendation algorithms have biases towards popular items.

In this paper, we presented re-ranking approaches for long-tail promotion and compared them to a state-of-the-art model-based approach. On two datasets, we were able to show that the re-ranking methods boost long-tail items while keeping the accuracy loss small, compared to the model-based technique. We also showed that the average recommendation popularity (ARP) measure from \cite{yin2012challenging} is not a good metric on its own for evaluating long-tail promotion, as algorithms might have similar ARP performance but quite different performance on other measures of popularity bias. So it is better to use it along with other metrics such as APLT and ACLT to get the right picture of the effectiveness of the algorithms.

One interesting area for future work would be using this model for multistakeholder recommendation where the system needs to make recommendations in the presence of different stakeholders providing the products \cite{umapHimanMS,burke_educational_2016,DBLP:conf/um/BurkeAMG16,abdollahpouri2019beyond}. In those cases, another parameter could be used to control the priority of each stakeholder in the system.


\bibliographystyle{ACM-Reference-Format}
\bibliography{main}

\end{document}